\documentclass[a4paper,11pt]{article}
\usepackage{pos}
\usepackage{dsfont}
\usepackage{mathtools}
\usepackage[T1]{fontenc}
\usepackage[utf8]{inputenc}

\usepackage[english]{babel}
\usepackage{a4, color}
\usepackage{amsmath}
\usepackage{siunitx}
\usepackage{lipsum}
\usepackage{chngcntr}
\usepackage{amsthm}
\usepackage[compat=1.1.0]{tikz-feynman}
\tikzfeynmanset{warn luatex=false}
\usepackage{braket}

\usepackage{comment}

\usepackage{slashed}
\usepackage{xcolor}
\definecolor{RED}{named}{red}
\usepackage{float}
\usetikzlibrary{matrix,fit,positioning}

\definecolor{energy}{HTML}{D81B60}   
\definecolor{eflux}{HTML}{1E88E5}    
\definecolor{mdens}{HTML}{43A047}    
\definecolor{press}{HTML}{FB8C00}    
\definecolor{shear}{HTML}{8E24AA}    
\usepackage{bm}

\title{Axion domain walls and thermal friction\footnote{Submitted to Proceedings of Science (PoS).}
}

\author*[a,b]{Amedeo M. Favitta}

\affiliation[a]{Dipartimento di Fisica e Chimica—Emilio Segrè, Università degli Studi di Palermo, \\Via Archirafi 36, I-90123 Palermo, Italy}

\affiliation[b]{INFN, Laboratori Nazionali del Sud,\\ Via S. Sofia 62, I-95123 Catania, Italy }

\emailAdd{amedeomaria.favitta@unipa.it}

\abstract{
The post-inflationary Peccei-Quinn symmetry-breaking scenario provides a rich theoretical
framework to study axion dark matter production through the dynamics of topological defects.
Accurate predictions for the axion abundance require a detailed understanding of the formation and evolution of cosmic strings and domain walls, which are inevitably produced in this scenario.

Most existing studies rely on large-scale numerical simulations of the classical equations of motion, which are subject to significant systematic uncertainties. In this contribution, we briefly
review the main features of the post-inflationary scenario and the current limitations in the literature, and
present preliminary results from a new analytical framework for axion domain wall networks.

Our approach is based on nonequilibrium quantum field theory. It employs the two-particle-irreducible effective action to derive effective Fokker-Planck equations for macroscopic quantities such as the average energy density and root-mean-square velocity of the network.We discuss the relevance
of thermal, self-interactions,  plasma-induced and quantum effects for cosmological axion abundance estimates, with
applications to QCD axions and high-mass axion-like particles.
}

\FullConference{
3rd Training School of the COST Action COSMIC WISPers (CA21106)\\
16--19 Sept 2025 \\
Annecy-le-Vieux, Campus du CNRS
}

\begin{document}
\maketitle

\section{Introduction}
Axions are prominent hypothetical particles originally proposed to solve the strong CP problem of QCD through the Peccei-Quinn mechanism
\cite{Peccei:1977ur,PhysRevLett.38.1440}.
The spontaneous breaking of a global $U_{\rm PQ}(1)$ symmetry gives rise to a Goldstone boson, later identified as the axion by Weinberg and Wilczek \cite{PhysRevLett.40.223,PhysRevLett.40.279}.
While early models with electroweak-scale symmetry breaking are excluded, "invisible axion"
models with symmetry-breaking scales above $10^{7}\,\mathrm{GeV}$ remain viable.

Despite their extremely weak interactions with the Standard Model, QCD axions provide an excellent
dark matter candidate, with an optimal mass range
$10^{-6}$-$10^{-2}\,\mathrm{eV}$ constrained by laboratory, astrophysical, and cosmological
observations
\cite{Sikivie:2008,DILUZIO202010,Favitta:2025neq}.
Similar considerations apply to axion-like particles (ALPs), which arise naturally in string theory compactifications, leading to the so-called axiverse \cite{Arvanitaki:2009fg}.

A central theoretical challenge arises in the post-inflationary Peccei-Quinn scenario, where the axion field is uncorrelated beyond the causal horizon and a network of topological defects forms in the Early Universe. Cosmic strings appear at the $U_{\rm PQ}(1)$ phase transition, while domain walls appear around the QCD crossover. The subsequent evolution and decay of this network
play a crucial role in determining the final axion abundance \cite{Sikivie:2008,Favitta:2025neq}.

While extensive numerical studies have been performed, various strategies lead to significantly different predictions for the QCD axion mass. These discrepancies motivate the
development of analytical approaches. In this work, we focus on domain wall networks, which are particularly sensitive to nonequilibrium effects.

\section{The model and theoretical framework}

We consider an effective low-energy description of axion dynamics in an expanding Universe,
based on the action \cite{Favitta:2025neq}
\begin{equation}
\mathcal{S}=\int d^4x\sqrt{|g|}
\left[
\frac{1}{2}\partial_\mu a\,\partial^\mu a
- V(a)
- \frac{1}{4}F^{\mu\nu}F_{\mu\nu}
- \frac{1}{4} g_{a\gamma\gamma} a\,\tilde F^{\mu\nu}F_{\mu\nu}
\right],
\end{equation}
with instanton potential with domain wall number $N_{\rm DW}$ and axion decay constant $f_a$
\[
V(a)=\frac{\eta^2 m_a^2}{N_{\rm DW}^2}
\left[1-\cos\!\left(N_{\rm DW}\frac{a}{\eta}\right)\right]=f_a^2 m_a^2
\left[1-\cos\!\left(\frac{a}{f_a}\right)\right]
\]
We work in a fixed flat FLRW background,
$ds^2=dt^2-R^2(t)\delta_{ij}dx^idx^j$.
The classical equations of motion admit domain wall solutions. In particular,
a more general collection of approximate solutions is the spherical ones, which can be written in the case of a center-of-mass moving in the $z$ direction as
\begin{equation}\label{dwsphere}
a(\vec x,t) \approx f_a\! \,\Big\{2\pi k
+4\arctan \exp{ \Big[m_a\,(r'-R) \Big]}\Big\},
\end{equation}
where $r'=\sqrt{(x-x_0)^2+(y-y_0)^2+\gamma^2(z-z_0-vt)^2}$ and $R$ is the radius of the sphere in its center-of-mass rest frame. 

As visible from the expression~(\ref{dwsphere}) in the laboratory frame, they are, in general, observed as ellipsoids which are Lorentz-contracted along the direction of motion.
A generalisation to a domain wall propagating in any spatial direction is straightforward.

To include quantum and thermal effects, we adopt a nonequilibrium quantum field theory framework
based on the Schwinger-Keldysh formalism and the two-particle-irreducible (2PI) effective action
\cite{Keldysh:1964ud,Favitta:2025neq}.
This yields coupled evolution equations for the mean field and its two-point function, from which effective transport equations can be derived.
The use of the Sine-Gordon potential provides a minimal and analytically tractable framework for studying axion domain wall dynamics.

\section{Effective Fokker-Planck description}
Assuming statistical isotropy of wall velocities and coarse-graining over microscopic degrees of
freedom, including the axion and plasma ones, the network dynamics can be described in terms of the average energy density $\rho$ and
the root-mean-square (rms) velocity $v$, in analogy with VOS models
\cite{Hindmarsh:1996xv,Martins:2016hbd}.
The resulting evolution equations read
\begin{align}
\dot{\rho} &=
- H \rho (1+3v^2)
- \frac{\rho}{l_f}
- \frac{c_w \rho}{L} v
+ P_{\rm lor}(T,v,\rho), \\
\dot{v} &=
(1-v^2)\left[
\frac{k_w(v)}{L}
- 3Hv
- \frac{\gamma v}{m_a l_f}
+ A_{\rm drag}(T,v,\rho)
\right].
\end{align}

The additional terms $A_{\rm drag}$ and $P_{\rm lor}$ arise from collisional interactions with the thermal
plasma and axion modes, and are expressed in terms of generalised collision integrals.
Unlike standard Boltzmann approaches, the collision operator is obtained by expanding the quantum
field around classical domain wall configurations rather than plane-wave modes.
This naturally incorporates thermal friction, decompression, and decoherence induced by axion and plasma
interactions \cite{Favitta:2025neq}.
\begin{figure}[h]
    \centering
    \includegraphics[width=0.8\linewidth]{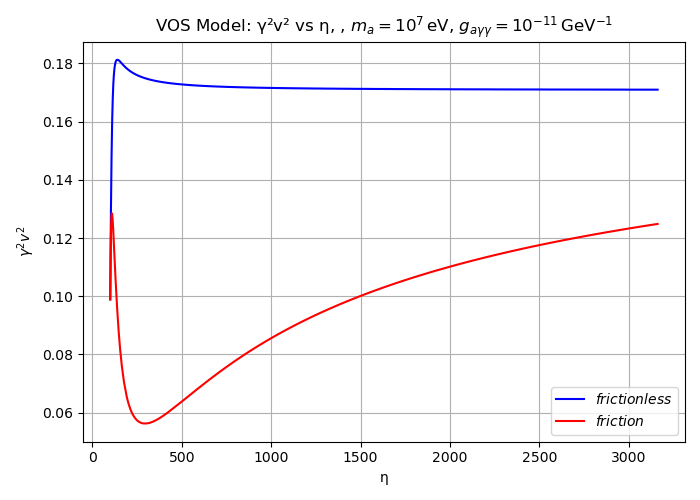}
\caption{Evolution of the rescaled rms velocity (right) of an axion
domain wall network as a function of conformal time $\eta$. We observe that the point of parameter space we consider here is far from the usual friction-dominated region of the literature (usually $m_a \gtrsim 10^{8}\,\rm{eV}$ and larger $g_{a \gamma \gamma}$). Figure from Ref.~\cite{Favitta:2025neq}, where we also show the dynamical evolution of the rescaled energy density $\rho \eta $.}
\label{fig:analytic}
\end{figure}
\section{Results}
Figure~\ref{fig:analytic} shows the evolution of the rescaled 
rms velocity $\gamma^2 v^2$ as a function of conformal time $\eta$.
Thermal friction leads to a suppression of wall velocities and modifies the approach to the
scaling regime.
The present analysis and our preliminary results demonstrate the usefulness of nonequilibrium quantum field theory for
better capturing thermal, self-interaction and plasma-induced effects in axion domain wall dynamics, motivating further investigation within this framework.
Extensions of this framework to more general settings will be considered in future work.

\acknowledgments{The author of the paper is grateful to Miguel Escudero, Amelia Drew, Pierre Sikivie, Roberto Passante, Lucia Rizzuto and Hyungjin Kim for valuable discussions and remarks. This article is based on work from COST Action COSMIC WISPers CA21106, supported by COST (European Cooperation in Science
and Technology).
}


\bibliographystyle{unsrt}
\bibliography{PoS}

\end{document}